\documentclass[twocolumn,amsmath,amssymb,prl,10pt,nofootinbib,superscriptaddress]{revtex4}
\usepackage{ graphicx, float,amsmath, amsmath, amssymb}
\usepackage[export]{adjustbox}

\def\be{\begin{equation}}
\def\ee{\end{equation}}
\def\bea{\begin{eqnarray}}
\def\eea{\end{eqnarray}}
\def\bse{\begin{subequations}}
\def\ese{\end{subequations}}

\usepackage[breaklinks, colorlinks, citecolor=blue]{hyperref}

\usepackage[labelsep=period]{caption}
\usepackage[normalem]{ulem}
\usepackage{soul}
\usepackage{minitoc}
\usepackage{physics}
\usepackage{caption}

\usepackage{bm}

\begin{document}

\title{Curvature bounce in general relativity: background and primordial spectrum}

\author{Cyril Renevey}%
\affiliation{%
Laboratoire de Physique Subatomique et de Cosmologie, Universit\'e Grenoble-Alpes, CNRS/IN2P3\\
53, avenue des Martyrs, 38026 Grenoble cedex, France
}
\affiliation{Institute for Theoretical Physics, ETH Zürich \\ Wolfgang-Pauli-Strasse 27, 8093 Zurich, Switzerland}

\author{Aurélien Barrau}
\affiliation{%
Laboratoire de Physique Subatomique et de Cosmologie, Universit\'e Grenoble-Alpes, CNRS/IN2P3\\
53, avenue des Martyrs, 38026 Grenoble cedex, France
}

\author{Killian Martineau}
\affiliation{%
Laboratoire de Physique Subatomique et de Cosmologie, Universit\'e Grenoble-Alpes, CNRS/IN2P3\\
53, avenue des Martyrs, 38026 Grenoble cedex, France
}
\author{Selim Touati}
\affiliation{%
Laboratoire de Physique Subatomique et de Cosmologie, Universit\'e Grenoble-Alpes, CNRS/IN2P3\\
53, avenue des Martyrs, 38026 Grenoble cedex, France
}

\date{\today}

\begin{abstract} 
Recent data suggest that the Universe could be positively curved. Combined with an inflationary stage, this might lead to a curvature bounce instead of the Big Bang. The background evolution is presented, as a function of the parameters controlling the cosmic evolution. The primordial tensor spectrum is also calculated and possible observational footprints of the model are underlined.  Several potentials are considered and general remarks are made about ``naturalness" in this context. 
\end{abstract}

\maketitle

\section{Introduction}

It is often argued that, unless radically new physics is used, the classical origin of the Universe is a singularity. This is not rigorously true. Following arguments given in \cite{Martin:2003bp,Uzan:2003nk,Barrau:2020nek}, we show that a curvature bounce might have occurred in the past. This does not require any exotic physics. The key ingredients of this scenario are the existence of an inflationary period -- which is obviously part of the standard cosmological model -- and a positive curvature for the spatial sections of the Universe. Although still actively debated and not fully consensual, a positive curvature seems to be favored by cosmological microwave background (CMB) measurements. If not truly established, this  hypothesis is anyway compatible with current data and this constitutes the core of this work. The idea of a ``curvature bounce" is not new but deserves a fresh look, taking into account the latest analyses.\\

This obviously raises some ``fine-tuning" issues. Although meaningful, those considerations are extremely difficult to formalize rigorously, due to the ambiguity associated with any arbitrary chosen measure. In this article, we do not address in details the naturalness of the model, instead we try to evolve backward in time the current state of the Universe and investigate the resulting history.  \\

Many different bouncing models have been considered in cosmology (see \cite{Battefeld:2014uga,Lilley:2015ksa,Brandenberger:2016vhg} for recent reviews). In particular, the bounce can be associated with an inflationary phase (see, {\it e.g.},  \cite{Piao:2003zm,Cai:2008qb,Liu:2013kea,Wan:2015hya,Ni:2017jxw,Bramberger:2019oss,Anabalon:2019equ,Ashtekar:2009mm,Martineau:2017sti}). This leads to a wide phenomenology depending on the detailed physical processes involved in the scenarii. This work focuses on the specific case were {\it no} new physics in involved, which  possible thanks to the curvature.\\

The consequences of a positive curvature are profound for the history of the Universe and might change the way we understand its ``origin". 
We first explain what are the motivations for a positive curvature. We then study in details the background behavior. The primordial tensor power spectrum is finally calculated. Throughout all the article we use Planck units.

\section{On the curvature of space}

The question of the spatial curvature of the Universe is an old one. It has been debated for decades. Although a flat space is usually considered as one of the key prediction of inflation, the actual situation is not that simple. Obviously, whatever the ``initial" curvature, inflation will make it negligible at the reheating.  However, as the Universe expands, the curvature contribution to the Friedmann equation will decrease more slowly ($\propto a^{-2}$) than the matter ($\propto a^{-3}$)  and radiation ($\propto a^{-4}$) ones. Curvature will then eventually dominate in the future, unless -- as it is the case in our Universe -- the cosmological constant ($\propto a^{0}$) overwhelms everything before this happens. 

Otherwise stated: the end of inflation might be the moment in cosmic history when the curvature contribution to the dynamics of the Universe is the smallest one. 

As recalled above, whatever its non-vanishing value, the relative contribution of the curvature increases as time goes on after inflation. But, importantly, it also increases when going backward in time during inflation: while the scalar field density remains roughly constant in the quasi-de Sitter stage, the curvature contribution increases (as it still scales as $\propto a^{-2}$) when the scale factor decreases. This might trigger a bounce instead of the singularity.\\

This question becomes especially important in the current context. First, the Planck data suggests that, using the CMB alone, the Universe might be positively curved  \cite{Aghanim:2018eyx}. The statistical significance of this result is weak. However, in addition, recent analyses presented in \cite{Park:2017xbl,Handley:2019tkm,DiValentino:2019qzk} considerably strengthen this possibility. In particular, it is shown in \cite{DiValentino:2019qzk} that the enhanced lensing amplitude in primordial power spectra, when compared to the prediction of the standard $\Lambda$CDM model, can be explained by this effect. This would also remove the tension within the Planck data about the values of cosmological parameters measured at different angular scales. The study concludes that Planck data favor a closed Universe with a probability of nearly 99.99\%.\\

Counter-arguments where given in \cite{Efstathiou:2020wem}. The actual conclusion heavily depends on the priors used for the analysis. At this stage, it is fair to say that a positive curvature is not firmly established but is worth being seriously considered. First, because some studies -- in particular using the CMB alone -- point in this direction and, second, because theoretical arguments -- in particular grounded in quantum gravity -- favor this hypothesis.

\section{Background behavior of a closed Universe}

We assume a homogeneous and isotropic background with closed spatial sections. The topology is $\mathbb{R}\times\mathbb{S}^3$, where $\mathbb{S}^3$ represents a hypersphere. The spatial curvature parameter $K>0$ is related to the physical radius $r(t)$ of the 3-sphere by $r^2(t)=a^2(t)/K$, where $a(t)$ is the dimensionless scale factor. Under these assumptions, the FLRW metric can be written as
\begin{align}
    \dd s^2=-\dd t^2+\frac{a^2(t)}{K}\left(\dd\chi^2+\sin^2(\chi) \dd\Omega^2\right).
\end{align}
The matter content of the Universe is represented by a perfect fluid of density $\rho$ and pressure $p$, such that the Einstein field equations lead to the Friedmann and Raychaudhury equations
\begin{align}
    H^2&=\frac{8\pi}{3}\rho-\frac{K}{a^2},\label{eq:friedmann}\\
    \dot{H}&=-4\pi(\rho+p)+\frac{K}{a^2},\label{eq:raychaudhury}
\end{align}
where $H=\dot{a}/a$ is the usual Hubble parameter. During the inflationary period, the matter content of the Universe can be described by a scalar field. In this work, we assume the inflaton to be a massive scalar field, that is to be described by the potential $V=m^2\phi^2/2$. In principle, it would make sense to consider other potentials -- a few remarks will be made on this point in the last section -- but the massive case makes the comparison with other models easier and captures most of the relevant phenomenology.  The Klein-Gordon equation in the expanding (or contracting) Universe reads
\begin{align}
    \ddot{\phi}+3H\dot{\phi}+\pdv{V(\phi)}{\phi}=0.\label{eq:klein-gordon}
\end{align}
The density and pressure of the inflaton field can be written as
\begin{align}
    \rho=\frac{1}{2}\dot{\phi}^2+V(\phi)\quad \textrm{and}\quad p=\frac{1}{2}\dot{\phi}^2-V(\phi).\label{eq:rho_p}
\end{align}
As well known, Eqs. \eqref{eq:friedmann}, \eqref{eq:raychaudhury} and \eqref{eq:klein-gordon} are not independent and only two of them are necessary to solve the system.\\ 

Let us first state that, when going backward in time, if the curvature density dominates the dynamics, while the Universe was in a de Sitter phase (or nearly so), a bounce inevitably takes place, as shown by the trivial analytical solution of the equations of motion:
\begin{equation}
a=\sqrt{\frac{3K}{\Lambda_i}}{\rm cosh}\left(\sqrt{\frac{\Lambda_i}{3}}t\right),
\end{equation}
where $\Lambda_i=8\pi\rho_{vac}$, with $\rho_{vac}$ the vacuum-like energy density of the field. Among bouncing models \cite{Battefeld:2014uga,Lilley:2015ksa,Brandenberger:2016vhg}, this one is specific in the sense that it does not require any exotic physics. \\

Following \cite{Barrau:2020nek}, it can easily be shown that the number of inflationary e-folds between the bounce and the reheating  is here given by:
\begin{equation}
N\approx\frac{1}{2} {\rm ln} \left( \frac{\rho_{R,0}}{\rho_{K,0}} \left[ (1+z_{eq}) \frac{T_{RH}}{T_{eq}} \right]^2  \right),
\label{n}
\end{equation}
where $\rho_{R,0}$ and $\rho_{K,0}$ are respectively the current densities of radiation and curvature, $z_{eq}$ is the redshift at the equilibrium time, $T_{RH}$ is the reheating temperature and $T_{eq}$ is the equilibrium temperature. The sudden reheating approximation is obviously crude \cite{Garcia:2020eof} but sufficient for this study. This formula does {\it not} use any dynamical feature of inflation: it just counts the amount of contraction (when thinking backward in time) ``needed" for the curvature density to equal the scalar field density (with the opposite sign in the Friedmann equation), which ensures the vanishing of the Hubble parameter at some point. When normalizing the curvature density at the value suggested in \cite{DiValentino:2019qzk}, this leads to $N\approx 65$ for $T_{RH}\approx 10^{16}$ GeV.  The inflation energy scale cannot be much above, otherwise this would conflict with current data, in particular with the tensor-to-scalar ratio upper limit. On the other hand, the number of e-folds cannot be much below this value so as to solve the usual cosmological paradoxes\footnote{In principle, one could relax this constraint by just requiring the number of inflationary e-folds to be equal to the number of post-inflationary e-folds but, in this specific framework, this also conflicts with data if it is much below 65, as shown in \cite{Barrau:2020nek}.}. 
This basically means that those parameters are somehow fixed in this model, which is to be contrasted with the usual cosmological framework where no upper bound exists on the number of inflationary e-folds and where the lower bound ont the reheating temperature is extremely weak. 

 As a consequence, from the purely numerical viewpoint, the value of the curvature can therefore be tuned so that the number of e-folds fits Eq.~\eqref{n}, at the considered density: this will automatically account for the (possibly) observed curvature. We have checked that it is indeed possible to find an initial value for the field -- associated with a given energy density -- so that the duration of inflation predicted by the model coincides with the intrinsic inflationary dynamics.\\

In order to have a strong control on the behavior of the inflationary period, we set the initial conditions (IC) of the background close to the onset of inflation, when $w_0:=p_0/\rho_0\approx -1$ at $t_0$.\\

Interestingly, the amount of inflation predicted in this framework is both high enough to be compatible with the observational lower bound and small enough so that non-trivial effects associated with the bounce could be seen in the CMB. In most bouncing models, the duration of inflation is so high \cite{bl,Bolliet:2017czc,Martineau:2017sti} that subtle footprints of the bounce are deeply super-Hubble and therefore non-observable. In this model, one might expect non-trivial features at large, but sub-Hubble, physical scales, that is in the low-$k$ part of the primordial spectra.\\

\begin{figure}
    \centering
    \includegraphics[width=8cm]{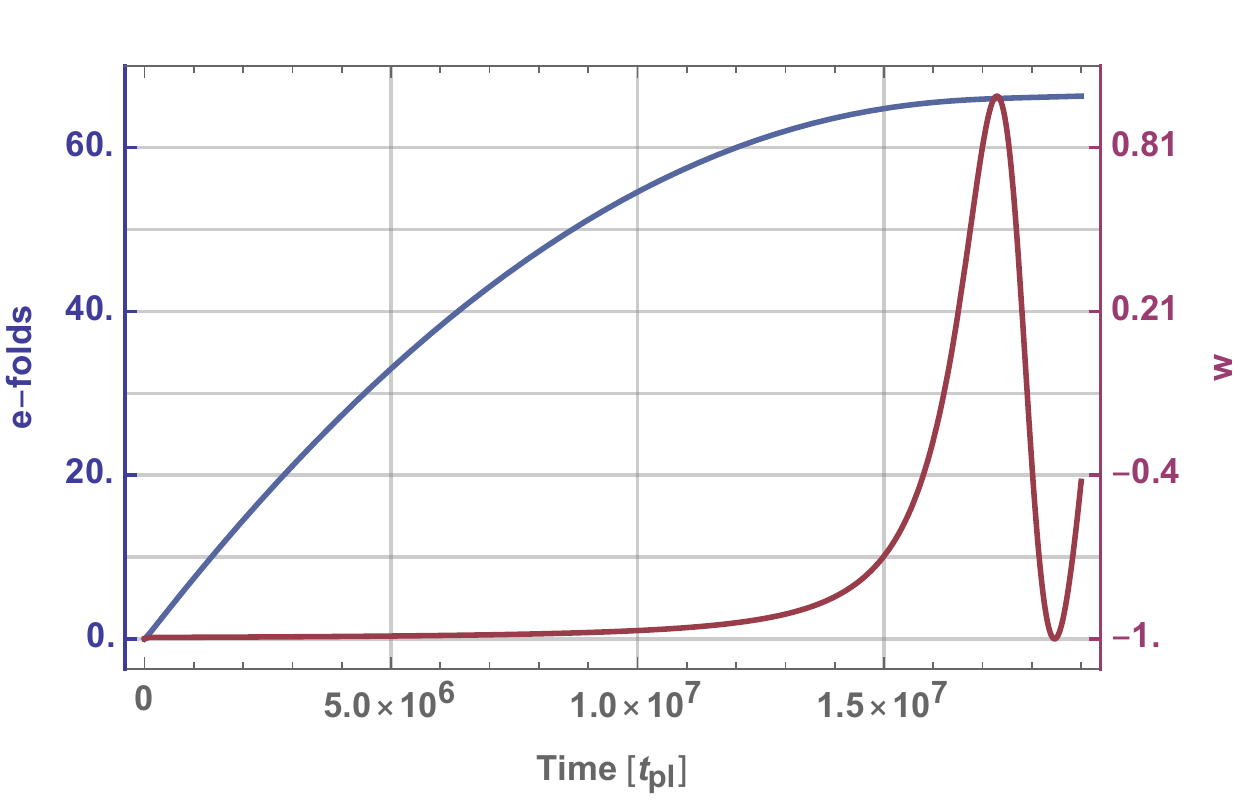}
    \captionof{figure}{{\it In blue:} Scale factor (variation measured in e-folds) during the inflationary period; {\it In purple:} $w=p/\rho$ during the same period. The density at $t_0$ has been chosen such that the number of e-folds during inflation is $N\approx 65$.}
    \label{fig:a_w_inflation}
\end{figure}

In the case of a massive scalar field, the number of e-folds can easily be analytically calculated and is given by $N=2\pi\phi^2-1/2$ in flat space. As the effect of the curvature is only relevant very close to the bounce, this result can also be safely  used  in a closed universe at the level of accuracy required here. At $t_0$, the onset of inflation, $N$ is directly connected to the density of the scalar field, using Eq.
~\eqref{eq:rho_p}. In order to have $N\approx 65$, as required for consistency, we choose $\rho_0=7.3\times 10^{-12}$.  Finally, we normalize the scale factor so that $a(t_0):=a_0=1$. Figure \ref{fig:a_w_inflation} shows the usual inflationary behavior, with $K=0$. Expanding the timeline to earlier times, $t<t_0$, would exhibit a singularity, where the scale factor tends to zero. At the time of reheating $t_{RH}$, the parameter $w$ starts to oscillate, as expected. 

A positively curved Universe filled with an inflaton field can avoid the Big Bang singularity if the curvature density $\rho_K:=3K/(8\pi a^2)$ is strong enough to compensate for the scalar field density at some point. This is basically due to the form of the Friedmann equation \eqref{eq:friedmann}, where the curvature density appears with the opposite sign (for a positive curvature) than the field density $\rho$. Furthermore, during inflation one has $w\approx -1$, which means that matter density is evolving as $\rho\propto a^{0}$ whereas $\rho_K\propto a^{-2}$. Hence, there should be have been a time in the past when $H^2=\frac{8\pi}{3}(\rho-\rho_K)$ vanishes, which represents a bounce. However, close to the singularity, the parameter $w$ is rising when going backward in time and if its value reaches $w= -1/3$ before both densities cancel each other, the Big Bang scenario is inevitable. This is analytically obvious as, if $w\geq -1/3$, the inflaton density varies faster than the curvature one.  If the curvature density is high enough at $t_0$, there will be a time in the past $t_B<t_0$, where $\rho_K(t_B)=\rho$.\\

The strategy of this study should now be clear: we basically fix a positive curvature for the spatial sections of the Universe and "impose" (as suggested by countless arguments) the existence of an inflationary stage. When evolving the equations of motion toward the past, a bounce takes place. Of course, if one thinks from the state of the Universe in the contracting branch, it is well known that the bounce will {\it not} be dynamically favored. The strong potential energy domination required for the (contracting) quasi-de Sitter stage to emerge, which is itself required for the bounce, is possible only for a small fraction of the phase space. We believe that two different questions should be disentangled. One is to figure out what the past of the Universe was. Taking into account what we know, the answer provided here makes sense ans strongly suggest the existence of a bounce. Another one is to try to understand why things happened in this way. We do not address here this second question which is extremely complicated and often ill-defined, in particular because of problems appearing when trying to deal with a reliable measure in cosmology. This issue is discussed in \cite{Barrau:2020nek}. We anyway stress here that, whatever its probability, the bouncing trajectory is indeed possible and even likely, if we think backward in time from our current knowledge. The instability with which it is associated when thinking forward in time \cite{Page:1984qt} is a very relevant question, but distinct from the viewpoint adopted here.\\

To ivestigate the likelihood of a bouncing trajectory when going backward in time, we consider the interval of $K$ values leading to this scenario. A first upper limit for the curvature comes from the consistency of the Friedmann equation. Indeed, since the left-hand side of Eq.~\eqref{eq:friedmann} is squared, the right-hand side must be positive. Taking this condition at the time $t_0$ leads to the requirement
\begin{align}
    K\leq \frac{8\pi}{3}\rho_0a_0^2\approx 6.1\times 10^{-11}.\label{eq:K_upper_bound}
\end{align}
The lower bound to the curvature parameter is set by requiring the existence of a bounce instead of the usual singularity. As the non-linearity of the equations of motion makes the analytical study of $w$ with respect to $a$ tedious, we have turned to numerical computations to determine the bound. The numerical study leads to the lower limit 
\begin{align}
    1.9\times10^{-11}\lesssim K.\label{eq:K_lower_bound}
\end{align}
Importantly, the values given by Eqs. (\ref{eq:K_upper_bound}) and (\ref{eq:K_lower_bound}) are close to one another. This interval is only valid when we take $w(t_0)=-1$. If one sets $w>-1$, it is possible to generate a longer period of inflation, before the initial conditions, as we will see later on.\\

\begin{figure}
    \centering
    \includegraphics[width=8.6cm]{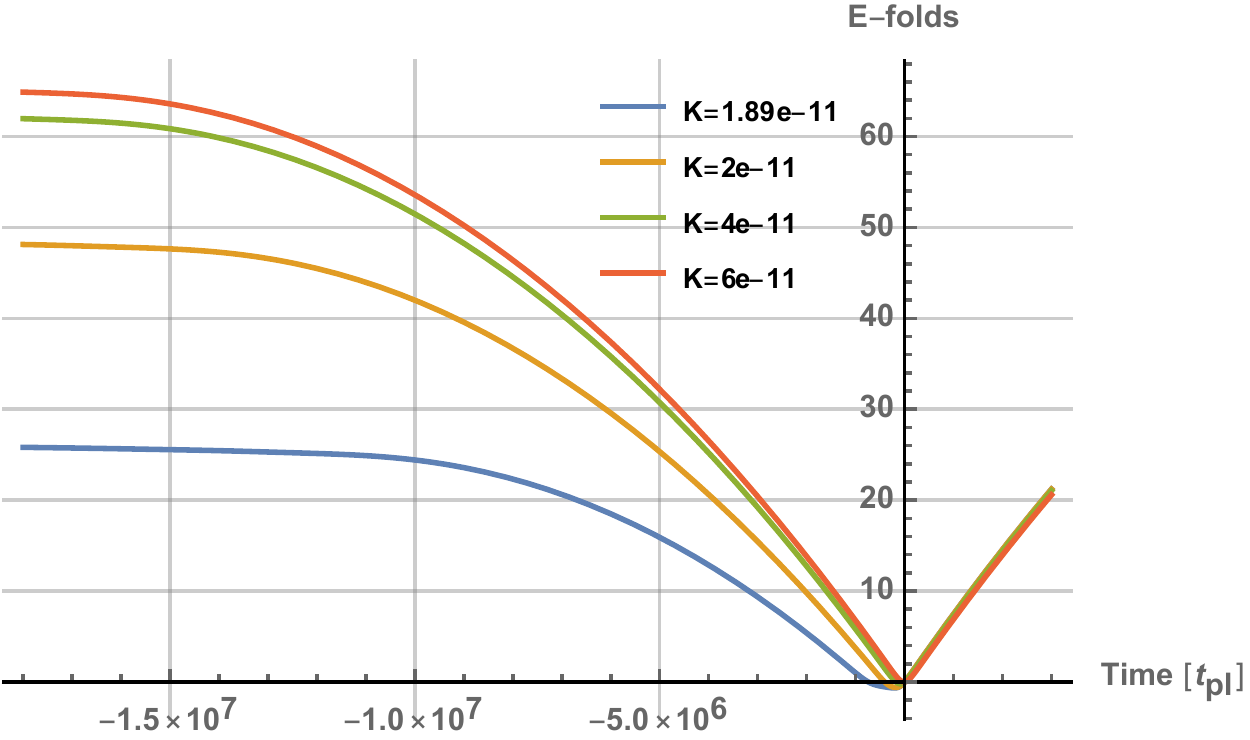}
    \captionof{figure}{Deflation behavior (e-folds) for different values of the curvature $K$.}
    \label{fig:duration_deflation}
\end{figure}

\begin{figure}
    \centering
    \includegraphics[width=8cm]{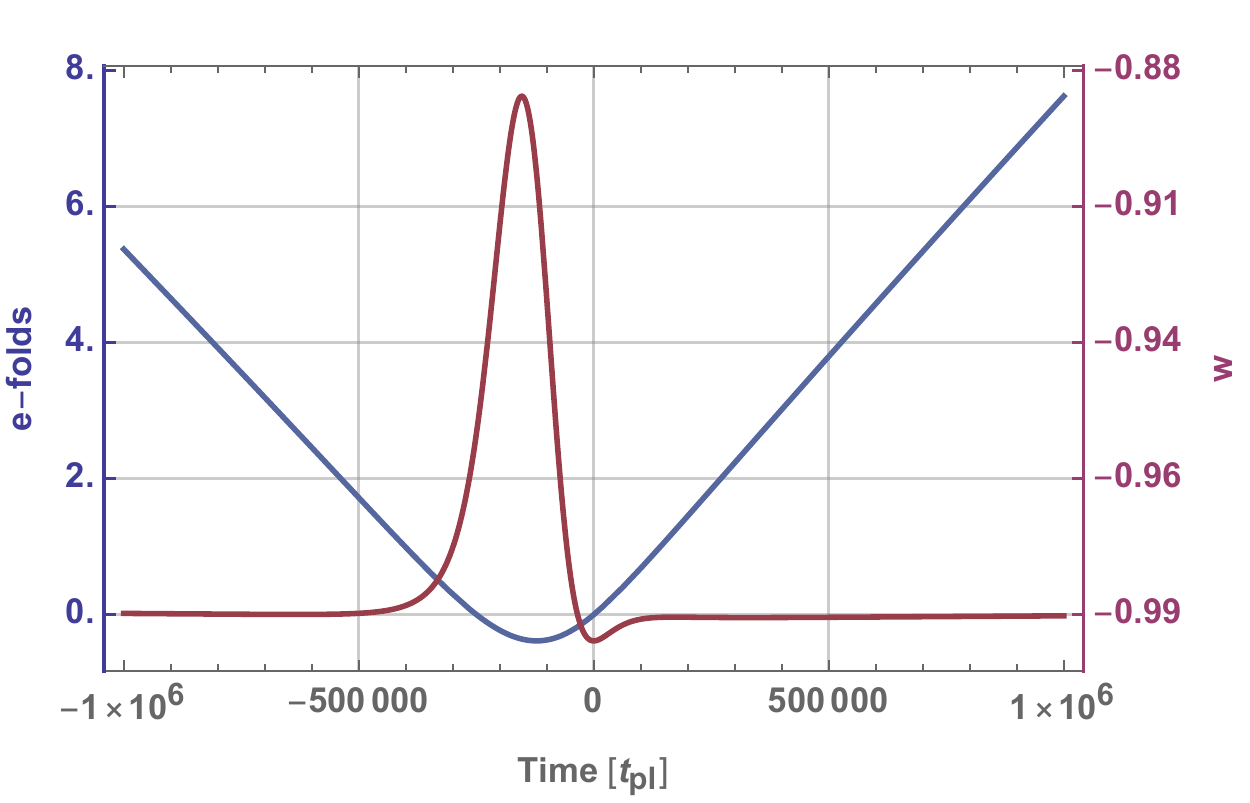}
    \captionof{figure}{Bouncing scenario with $K=3\times 10^{-11}$. {\it In blue:} Scale factor (variation measured in e-folds) close to the bounce; {\it In purple:} $w=p/\rho$.}
    \label{fig:bounce}
\end{figure}

\begin{figure}
    \centering
    \includegraphics[width=8cm]{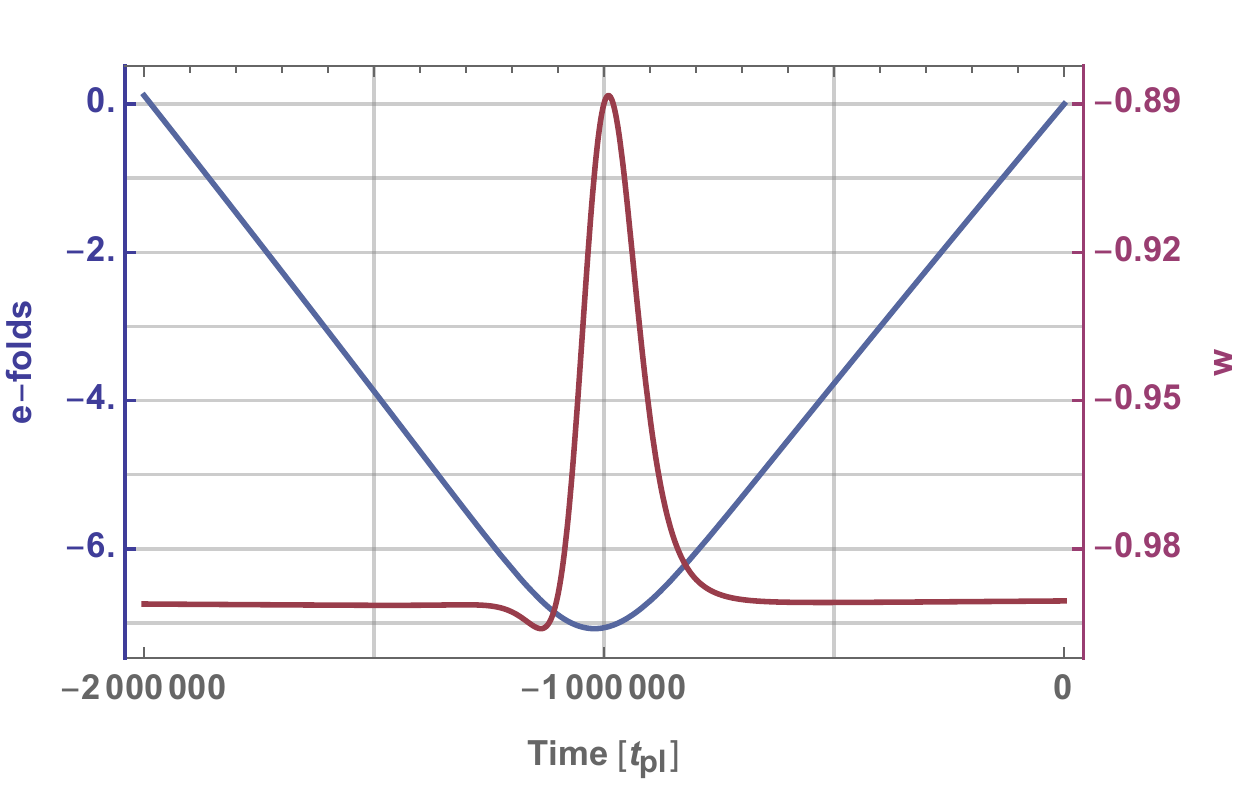}
    \captionof{figure}{Bouncing scenario with the experimentally valid curvature and inflation parameters. {\it In blue:} Scale factor (variation measured in e-folds) close to the bounce; {\it In purple:} $w=p/\rho$.}
    \label{fig:bonga_bounce}
\end{figure}

\begin{figure}
    \centering
    \includegraphics[width=8cm]{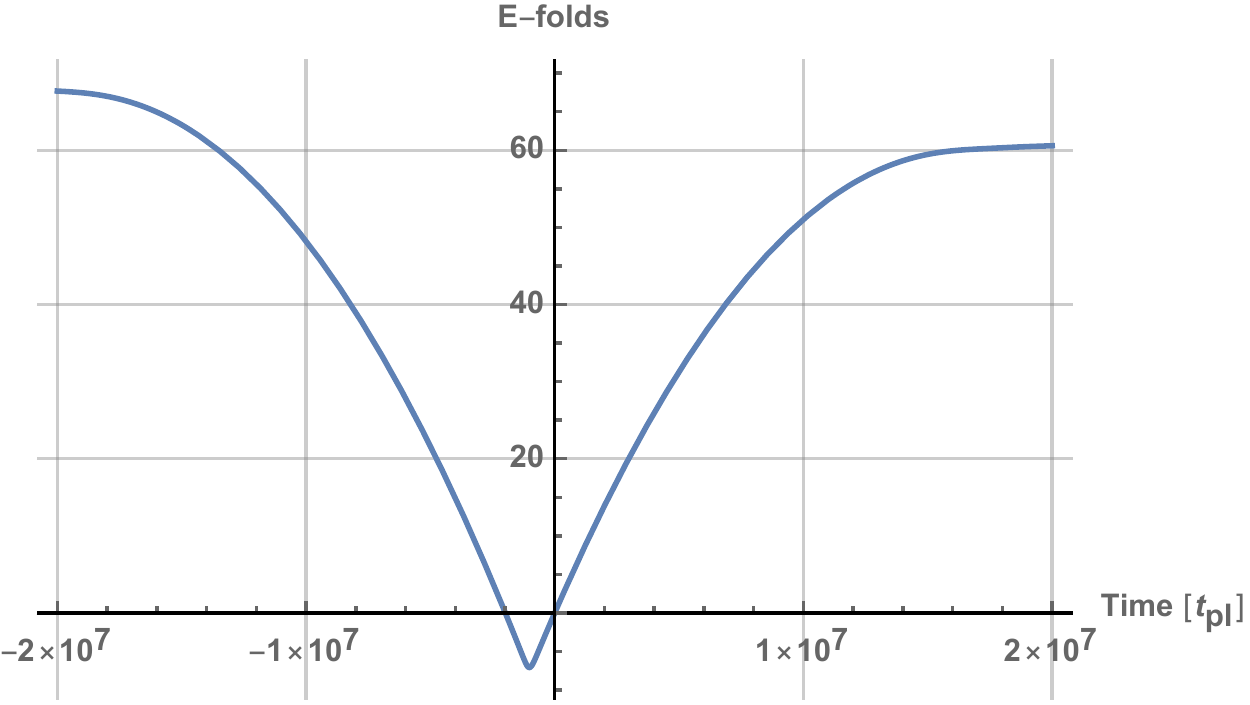}
    \captionof{figure}{Deflation and inflation with experimentally valid curvature and inflation parameters at time $t_\star=0$.}
    \label{fig:bonga_long}
\end{figure}

A range of possible values for $K$ leading to a bouncing scenario has therefore been found, when considering the initial conditions stated above. As expected, different values of $K$ lead to different bouncing solutions. In particular, the duration of the pre-bounce deflation is impacted by the strength of the curvature at the bounce. This is an important feature. In Fig.~\ref{fig:duration_deflation}, we have drawn the deflation period for different values of $K$. The weaker the curvature at the initial time $t_0=0$, the shorter the deflation period. 
This raises an interesting point. One of the appealing features of this bouncing scenario is to avoid the Big Bang singularity. However, if the deflation period was too brief, one will inevitably face another singularity {\it in the past}. The cosmological constant will indeed never dominate over the curvature density when going backward in time before the bounce. Another bounce (at low density) will take place, leading to a contraction -- still thinking backward in time -- and to a singularity when going further to the past. This might be evaded by selecting very specific conditions, but the argument cannot be meaningfully repeated for each contracting phase. A long enough deflation is therefore mandatory  so that the cosmological constant protects the Universe from a re-contraction in the remote past. Figure~\ref{fig:duration_deflation} shows that this is possible. There also exist scenarios where the period of deflation is longer than the period of inflation. One can indeed notice on Fig.
~\ref{fig:bounce} that the behavior of $w$ is not symmetric around the bounce, which causes the difference in duration between deflation and inflation. If one sets $w$ slightly over $-1$ at the initial time $t_0$, it is possible to flip the behavior of $w$ and to obtain a longer period of deflation. This effect can be seen in Figs. \ref{fig:bonga_bounce} and \ref{fig:bonga_long}.\\

It is obviously important to check that the allowed interval for $K$, corresponding to a bouncing scenario, is consistent with experimental data. Having a concrete answer to this question is however difficult. Since the experimental curvature is measured in the contemporary Universe, one needs to know the number of e-folds from the beginning of inflation to the present time to set the initial value which, itself, influences the dynamics. Some specific initial conditions were suggested in \cite{Bonga:2016iuf,Bonga:2016cje}, where the effect of curvature during inflation was also studied (but ignoring the presence of a bounce). In particular, the time $t_\star$ at which the modes $k=0.05\,$Mpc$
^{-1}$ exited the horizon is used to set the initial conditions. If one normalizes the scale factor at the time $t_\star$ and uses $\Omega_{k,0}=-0.044$ \cite{DiValentino:2019qzk}, the resulting curvature is $\bar{K}\approx 5\times 10^{-17}$. The other parameters, such as the density and $w$, are also different from what we used. Even though the density is slightly less than what was used earlier, leading to a number of e-folds $N\approx 60$, the experimental curvature $\bar{K}$ is way lower than the range we previously calculated. However the initial conditions of \cite{Bonga:2016iuf,Bonga:2016cje} are not set at the onset of inflation, but at a later time when $w>-1$. Due to the difference in evolution between $\rho$ and $\rho_K$, there is no inconsistency. We thus rely on numerical computations to check the plausibility of the bouncing scenario. Simulating a long period of inflation when going backward in time is unstable, even in flat space. Very fine-tuned initial conditions can lead to a long period of inflation in the past, but in most cases, inflation stops abruptly in a singularity. Varying the values of $\phi_\star$ and $\phi'_\star$ given in \cite{Bonga:2016iuf,Bonga:2016cje} within a $1\%$ interval does indeed lead to bouncing scenarios with a long enough inflation period, as shown in Figs. \ref{fig:bonga_bounce} and \ref{fig:bonga_long}.
However, as explained before and as pointed out in \cite{Barrau:2020nek}, the interval considered here for $K$ is anyway compatible with data by construction of the model. It indeed leads to the amount of inflation precisely needed for the current curvature to be given by the value estimated in \cite{DiValentino:2019qzk}.

\section{Primordial power spectrum for tensor modes}

The calculation of perturbations in a closed Universe has been considered in many studies. Interesting results were recently derived, {\it e.g.} in \cite{deHaro:2015yyh,Asgari:2018wio,Akama:2018cqv}, to cite only a few. We shall here focus specifically on the approach developed in \cite{Bonga:2016iuf,Bonga:2016cje}. One considers a family of metrics
$g_{\mu\nu} (\varepsilon)$:
\begin{equation}
g_{\mu\nu}(\varepsilon) = a^2(\eta) \left( \bar{g}_{\mu\nu} + \varepsilon h_{\mu\nu} \right) \, ,
\end{equation}
where $a^2 \bar{g}_{\mu\nu}$ is the background metric, $h_{\mu\nu}$ is a linear perturbation, $\varepsilon$ is a small parameter, and $\eta$ is the conformal time. The relevant spatial components read:
\begin{equation}
\label{eq:eomtensor}
\bar{\grad}^2 h_{ij}^{TT} - 2 \frac{a'}{a} h_{ij}^{' \, TT} - \frac{2}{r_o^2} h_{ij}^{TT} = 0 \, ,
\end{equation}
where the prime denotes here a derivative with respect to $\eta$, $r_o^2 = 1/K$, and $i,j=1,2,3$. Instead of expanding the equation of motion in  Fourier space, as usually done, one expands the perturbations on the tensor hyperspherical harmonics 
$\mathcal{Q}^{nlm,s}_{ij}(\chi,\theta,\phi)$ \cite{Gerlach:1978gy}. They are eigenfunctions of the Laplacian operator $D^2$:
\begin{equation}
D^2 \mathcal{Q}^{nlm,s}_{ij}(\chi,\theta,\phi)=-\frac{(n^2-3)}{r_o^2} \mathcal{Q}^{nlm,s}_{ij}(\chi,\theta,\phi) \,.
\end{equation}
Importantly, the factor $(n^2-3)/r_o^2$ plays a role equivalent to the one of the wavenumber squared $k^2$ in flat space \cite{Lewis:1999bs}. Rescaling the coefficients of the development of $h$ (we skip here the integer labels for simplicity) by $\mu=a(\eta) h$ and promoting the resulting field to be an operator, one is led for the coefficients of the creation and annihilation operators to:
\begin{equation}
e_n^{''}  + \left(\frac{n^2 -1}{r_o^2} - \frac{a^{''}}{a}\right) e_n = 0 \,,\label{eq:operator_equation}
\end{equation}
with the normalization
\begin{equation}
 e e^{\star\prime} - e^\prime e^\star= i~.\label{eq:operator_normalization}
\end{equation}
The main effects of curvature are therefore twofolds. First, the curvature obviously changes the background evolution which leaves a footprint on the perturbations. Second, it discretizes the effective wavenumber.\\

Setting initial conditions for the perturbations is a key question when evaluating power spectra. The problem is known for being very difficult in bouncing models. In particular the case of scalar perturbations is problematic because the term $z''(\eta)/z(\eta)$ entering the harmonic oscillator equation does {\it not} generically tends to zero in the remote past (see, {\it e.g.}, \cite{Barrau:2018gyz}). This means that it is hard to disentangle effects specifically due to the bounce from effects associated with the lack of a unique privileged vacuum. In addition, the equation of motion is usually very complicated and the relation between $z$ and $a$ is non-trivial and depends on the potential. For those reasons,  we focus here on tensor modes. \\

In a bouncing scenario, there is {\it stricto sensu} no possibility to define an ``initial time". In the specific context of loop quantum cosmology, the point of view that the bounce time should be chosen to set initial conditions was advocated, {\it e.g.}, in \cite{Agullo1}, whereas the opposite vision was proposed, {\it e.g.}, in \cite{Schander:2015eja}.

\begin{figure}
    \centering
    \includegraphics[width=8cm]{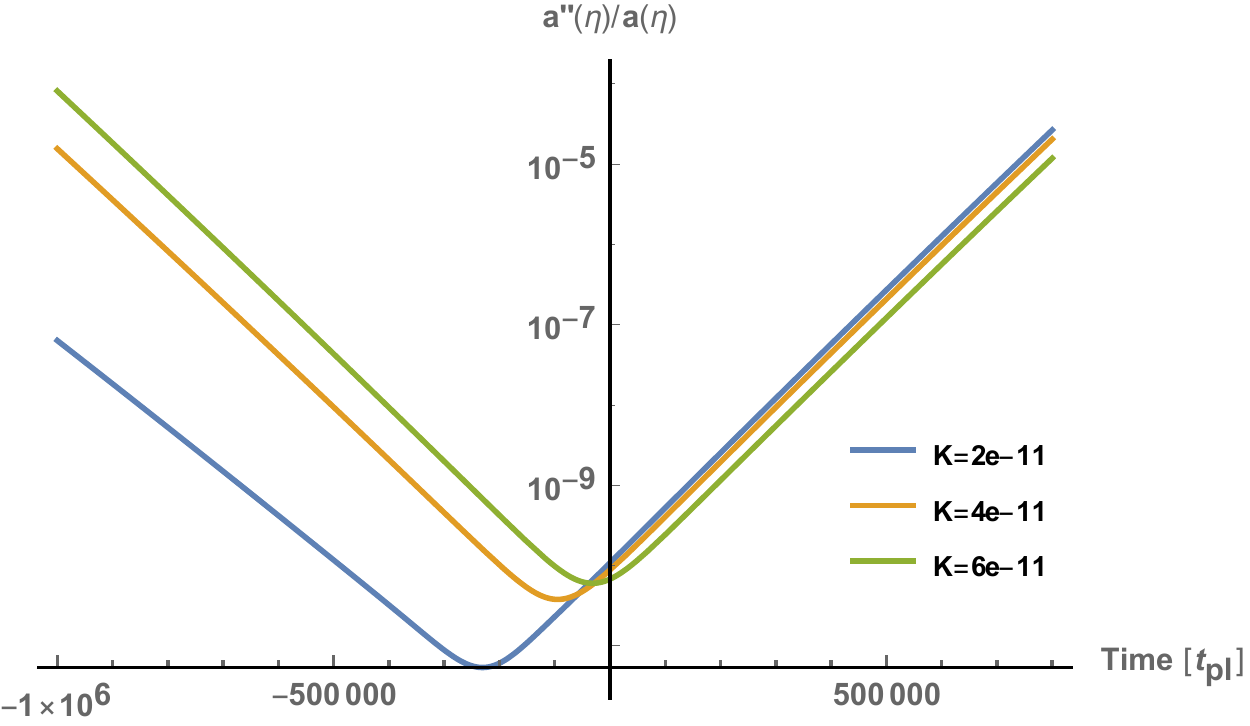}
    \captionof{figure}{Potential of the tensor modes for different values of the curvature.}
    \label{fig:z_potential}
\end{figure}

In the set-up considered here, there is necessarily a deflation stage before the bounce. This is obviously necessary for the bounce to take place and this is anyway necessary for the curvature to dominate: without deflation, any amount of (pre-bounce) matter or radiation density would grow faster than curvature. This deflation stage makes it impossible to set initial conditions in the past as $a''(\eta)/a(\eta)$ does not anymore decrease when going backward in time before the bounce and no Bunch-Davis-like vacuum is approached. However, the $a''(\eta)/a(\eta)$ term is very small close to the bounce and this selects, at least at the heuristic level, a preferred time to set initial conditions. The behavior of $a''(\eta)/a(\eta)$ is shown for different values of $K$ in Fig. \ref{fig:z_potential}.\\

Considering Eq.~\eqref{eq:operator_equation}, one can recognize a harmonic oscillator with a time-dependent frequency
$$\omega_n^2(\eta)=\frac{
n^2-1}{r_o^2}-\frac{a''(\eta)}{a(\eta)}:=k_\textrm{eff}(n)-\frac{a''(\eta)}{a(\eta)}.$$
In our case, since $r_o^2\sim 10^{10}$ and $a''(t_B)/a(t_B)\sim 10^{-10}$, one cannot use a rigorous Minkowski vacuum $k_\textrm{eff}\gg a''/a$ at the initial time $t_B$ for small discret wavenumbers $n$. As in \cite{Bonga:2016cje}, we choose an instantaneous vacuum state that minimizes the Hamiltonian at the initial time. In order for waves to propagate with a positive frequency and using the normalization of Eq. \eqref{eq:operator_normalization}, we can set the basic mode functions at $t_B$ as
\begin{align}
    e_n(t_B)&=\frac{1}{\sqrt{2\omega_n(t_B)}},\\
    e_n'(t_B)&=-i\sqrt{\frac{\omega_n(t_B)}{2}}.
\end{align}\\

\begin{figure}
    \centering
    \includegraphics[width=8.5cm]{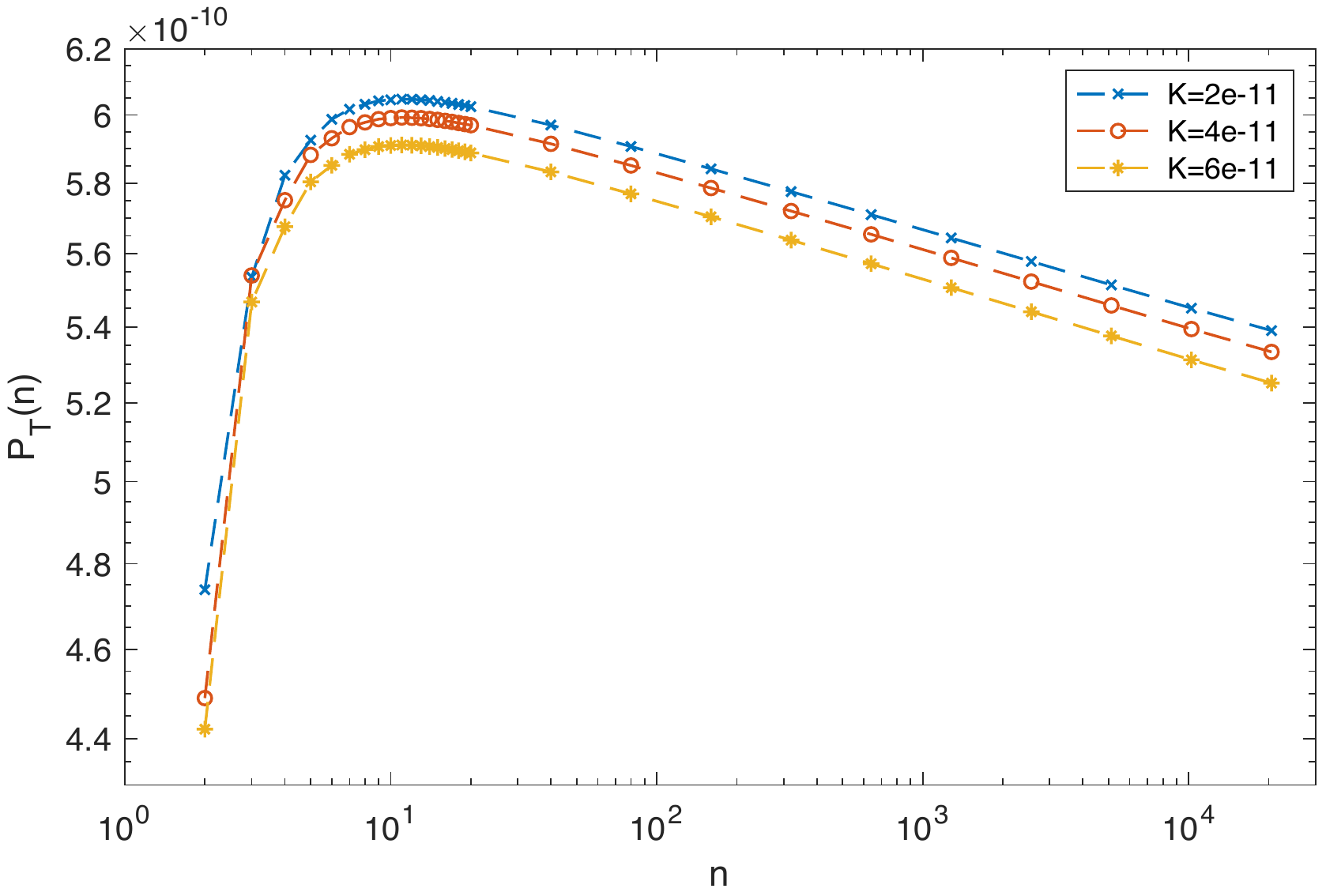}
    \captionof{figure}{Primordial power spectrum for the tensor modes for different values of the curvature. The discrete wavenumber $n$ has been calculated with $k^2=(n^2-1)\,2\times 10^{-11}$.}
    \label{fig:spectrum}
\end{figure}

In order to compute the primordial power spectrum for tensor modes $\mathcal{P}_T(n)$, we will proceed in the usual way. Starting from the initial time $t_B$, we simulate the mode functions $e_n(t)$ up to $t_e$, where $k_\textrm{eff}=a(t_e)H(t_e)$. The primordial power spectrum can then be calculated \cite{Martineau:2019kuc}:
\begin{align}
\mathcal{P}_T(n)=\left.\frac{32k_\textrm{eff}^3}{\pi}\abs{\frac{e_n}{a}}^2\right|_{t=t_e}.
\end{align}
The results of this analysis for different values of $K$ are shown in Fig. \ref{fig:spectrum}. The power spectra for different values of $K$ are nearly parallel for $n>20$. As the initial time when we set the initial conditions varies with respect to K, the amplitude of the power spectra differs slightly. More importantly, one can notice that the curvature significantly affects the large-scale modes. They are damped due to the effect of the closure of space, which is in concordance with the result of \cite{Bonga:2016cje}, where the bounce itself is however not considered.\\

It is interesting to try to figure out whether the non-trivial effects appearing in the low-$k_{eff}$ part of the power spectrum are observable in the cosmological microwave background (CMB). To this aim, one needs to relate the comoving wavenumber $k_c(t_r)\equiv k_{\textrm{eff}}(t_r)$ with its physical counterpart $k_p(t_r)$ at the time of recombination $t_r$. Using $N\approx 65$ and $T_{RH}\approx 10^{16}\,$GeV one can get the orders of magnitude. The physical wavenumber  observed in the CMB are of the order $\bar{k}_{p}\approx 0.05\,$Mpc$^{-1}\approx 3\times 10^{-59}$ and the scale factor at recombination can be estimated using $a(t_r)\approx e^N\times T_{RH}/T_{r}$, where $T_r\approx 0.3\,$eV \cite{Aghanim:2018eyx}. Straightforwardly, it follows that $\bar{k}_c\approx 2\times 10^{-5}$ and with a curvature constant of the order $K\sim 10
^{-11}$, we obtain the discrete comoving wavenumber of the CMB, $\bar{n}\approx 4$. This is just in the range where the dampening effect is observable. This result is however very sensitive to the number of e-folds. As the tensor modes have not yet been observed, we leave the accurate simulation for future studies. This work however shows that some features of the curvature bounce might be observable thanks to the limited number of inflationary e-folds inherent to this model.

\section{Curvature Bounce with different potentials}

The massive scalar field is a good toy model to study the inflationary period, but it is disfavored by recent data, relatively to other more realistic potentials, such as the Starobinsky one \cite{Akrami:2018odb,Starobinsky:2001xq}. It is therefore meaningful to study the cosmological background behavior in the presence of spatial curvature, as well as the likelihood of a bouncing scenario, with such a potential. In addition, a flat potential is expected to make the bounce more ``likely" when starting from the contracting phase. Another flat potential especially designed to this aim was also considered in \cite{Matsui:2019ygj}.\\

The Starobinsky potential is of the form 
\begin{align}
    V(\phi)=\frac{3M^2}{32\pi}\left(1-e^{-\sqrt{16\pi/3}\phi}\right)^2, \label{eq:starobinsky_potential}
\end{align}
where $M$ is a mass scale parameter. A realistic value of the parameter $M$ can be determined following the reasoning of \cite{Bonga:2015xna}, that is using the slow-roll parameters as well as the spectral index $n_s$ and the amplitude $A_s$. The potential \eqref{eq:starobinsky_potential} was also studied in \cite{Bonga:2016cje}, where no bouncing scenario was encountered. However, following the same strategy as we did before, {\it i.e.}
~varying the initial conditions for the scalar field at the initial time $t_\star$ in a small interval, one can generate a sufficiently long period of inflation prior to the initial time to have a curvature bounce. This shows, as expected, a high sensitivity to initial conditions. \\

It is extremely difficult to have a quantitative estimate for the ``probability of the bounce" for a given potential. It is well known (see, {\it e.g.}, the discussion in \cite{Bolliet:2017czc}) that depending on the chosen measure and selected surface for the initial conditions one can be led to completely opposite conclusions. This is a lively debate in cosmology \cite{Schiffrin:2012zf}. Nevertheless, it is possible to get a vague qualitative idea of the likelihood of the bounce scenario for the Starobinsky potential compared to the massive scalar field case. We set the initial conditions at time $t_0$, the onset of inflation, where $w=-1$ and the initial density $\rho_0$ is chosen such that we have $N\approx 65$ e-folds. In this new case, this means $\rho_0\approx 1.6\times 10^{-13}$. As before, one can find a range of possible spatial curvature $K$ such that a bounce takes place. The upper bound is found using \eqref{eq:K_upper_bound}, leading to $K<13\times 10
^{-13}$, and using numerical computations we find the lower bound $K> 1.1\times 10^{-13}$. Relatively to the density, the range of curvature values is indeed larger in the case of a Starobinsky potential than for the massive scalar field. It, however, remains quite narrow. The behavior of the background is similar to the previous case.\\

\begin{figure}
    \centering
    \includegraphics[width=8cm]{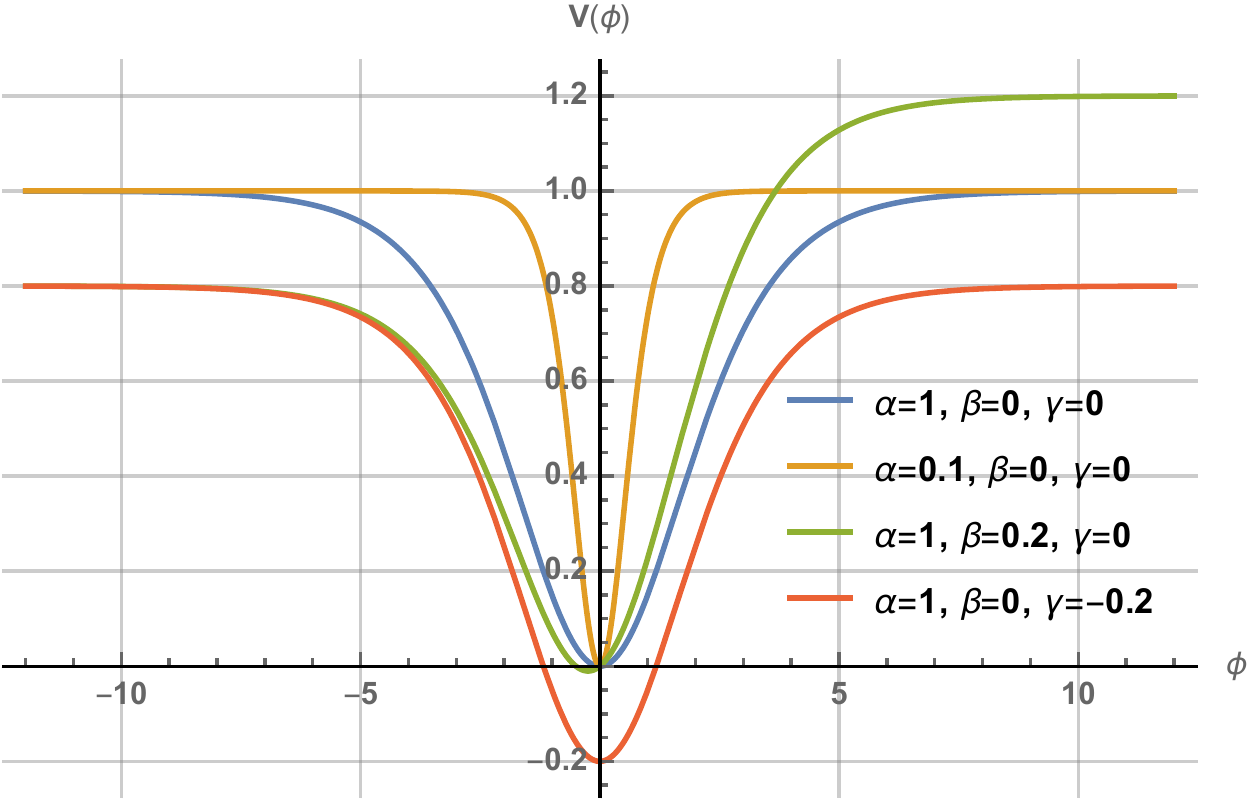}
    \captionof{figure}{Scalar potential \eqref{eq:matsui} for different values of the parameters $\alpha$, $\beta$ and $\gamma$, where we set $V_0=1$.}
    \label{fig:matsui}
\end{figure}

A potential of the form
\begin{align}
    V(\phi)=V_0\left(\tanh^2\left[\frac{\phi}{\sqrt{6\alpha}}\right]+\beta\tanh\left[\frac{\phi}{\sqrt{6\alpha}}\right]+\gamma\right),\label{eq:matsui}
\end{align}
where $V_0>0$, $\alpha>0$, $-1<\beta<1$ and $-1<\gamma\leq 0$, was proposed in \cite{Matsui:2019ygj} to make the bounce ``natural". The effect of varying the parameters is illustrated in Fig. \ref{fig:matsui}.
As one can see, $\alpha$ controls the steepness of the drop at $\phi=0$, $\beta$ changes the symmetry of the potential and $\gamma$ sets the amplitude. We have indeed checked that some values of the parameters, {\it e.g.} $\gamma=\beta=0$ and $V_0\approx K$, do indeed systematically ensure a bounce. This is an important way to circumvent the apparent fine-tuning of the model when thinking forward in time. In principle, if this bouncing scenario were established, this might even allow one to select some families of potentials compatible with a reasonable probability of occurence of the bounce.

\section{Conclusion}

In this article, we have shown that under the experimentally valid assumption of a positively curved space (and of  an inflationary stage in the past), a bounce might have naturally taken place. This basically fixes the number of e-folds -- which can be anything above 65 in standard cosmology -- and the reheating temperature. This removes the Big Bang singularity without any exotic physics. The view adopted here is the one of a historian who tries to figure out what was the past knowing what we know. We have not addressed the question of determining {\it why} the Universe might have followed this specific trajectory, which is a different question.\\

It should be noticed that the model is consistent in the sense that the density never approaches the Planck density. It is therefore legitimate to neglect quantum gravity effects. The physical size of the Universe also remains much larger than the Planck length, even at the bounce time. Both from the density and from the size points of view, the quantum geometry regime is never reached.\\

We have studied in details the way in which the background behavior depends on the contingent parameters and on the inflaton potential. We have also calculated the primordial tensor power spectrum and shown that footprints of the curvature bounce might be observable at large scales. We have checked that some potentials make the bounce apparently more ``natural".\\

Many open questions remain to be addressed in the future. The issue of fine-tuning and the backreaction effect of fluctuations should obviously be considered seriously. Both because of the measurement problem and because of possible anthropic considerations those points are clearly difficult to answer. The primordial scalar power spectrum should also be calculated. The main trend -- that is a suppression of power in the infrared part -- is expected to be the same than for tensor modes. This should make the agreement with data even better (although not significantly, due to the cosmic variance) than in the usual model. In the spirit of \cite{Gao:2014hea,Gao:2014eaa}, the question of non-gaussianities should be addressed. Our predictions should also be confronted to the ones obtained in \cite{Lilley:2011ag} for a symmetric bounce.

\bibliography{refs}

\end{document}